\documentclass[%
 reprint,
 prapplied,
 groupedaddress,
superscriptaddress,
 amsmath,amssymb,
 aps,
]{revtex4-2}
\usepackage{graphicx}% Include figure files
\usepackage{dcolumn}% Align table columns on decimal point
\usepackage{bm}% bold math
\usepackage{siunitx}
\usepackage{physics}
\usepackage[version=4]{mhchem}
\usepackage{txfonts}
\usepackage{hyperref}
\usepackage{hyperref}
\usepackage[version=4]{mhchem}
\hypersetup{
    colorlinks=true,
    linkcolor=blue,
    filecolor=magenta,
    urlcolor=blue,
    pdftitle={Widefield AC Susceptibility},
    pdfauthor={Dasika Shishir},
    }
\begin{document}

\preprint{}

\title{Real-time Simultaneous Dual Sensing of Temperature and Magnetic Field using NV-based Nano-diamonds}
\begin{abstract}
Quantum sensors based on Nitrogen Vacancy (NV) centers in diamond are highly capable of sensing multiple physical quantities. In this study, we use amplitude-modulated lock-in detection of optically detected magnetic resonance of NV nanodiamonds (NVND) to investigate the link between temperature (T) and the zero-field splitting parameter (D) and also the relationship between magnetic field values and the difference of resonance frequencies. We also present NVNDs' capacity to simultaneously sense both thermal and magnetic fields in real time. This dual-sensing approach is beneficial for studying magnetic materials whose magnetization depends on temperature and the applied magnetic field, such as certain ferromagnetic and ferrimagnetic materials. Integrating real-time thermal and magnetic field measurements provides unique opportunities for failure analysis in the integrated circuit (IC) industry and for studying thermodynamic processes in cell physiology. The ability to concurrently monitor temperature and magnetic field variations offers a powerful toolset for advancing precision diagnostics and monitoring in these fields.

\end{abstract}
\author{Sonia Sarkar}
\affiliation{Center for Research in Nano Technology and Science,
Indian Institute of Technology Bombay, Mumbai, Maharashtra, India}
\affiliation{Department of Electrical Engineering, Indian Institute of Technology Bombay, Mumbai, Maharashtra, India}
\email{soniasarkar@iitb.ac.in}
\author{Namita Agarwal}
\affiliation{Department of Physics, Indian Institute of
Technology Bombay, Mumbai, Maharashtra, India}
\author{Dasika Shishi}
\affiliation{Department of Electrical Engineering, Indian Institute of Technology Bombay, Mumbai, Maharashtra, India}
\author{Kasturi Saha}
\email{kasturis@ee.iitb.ac.in}
\affiliation{Department of Electrical Engineering, Indian Institute of Technology Bombay, Mumbai, Maharashtra, India}
\affiliation{Center of Excellence in Quantum Information, Computing Science and Technology, Indian Institute of Technology Bombay, Powai, Mumbai--400076, India}
\affiliation{Center of Excellence in Semiconductor Technologies (SemiX), Indian Institute of Technology Bombay, Powai, Mumbai--400076, India}

\maketitle

\section{Introduction}
\label{sec:introduction}
Real-time and simultaneous measurement of temperature and magnetic field is of great importance in many applications. For example, the magnetic field emanating from an electronic device is indicative of the current flowing through it. At the same time, a temperature measurement is necessitated by the fact the many properties of electronic devices are heavily dependent on the
temperature\,\cite{PhysRev.82.900,sze2021physics, Zhang2019}. Another area where simultaneous detection of magnetic field and temperature can be of great value is the study of magnetic materials. The stray magnetic field from the magnetic material is directly proportional to the magnetization of the sample, while the magnetization itself being a function of temperature\,\cite{Mugiraneza2022, Zhang2022}. Typically a ferro-magnet loses its magnetization beyond the Curie temperature\,\cite{cullity2011introduction}.
A wide variety of sensors exist for magnetic field and temperature sensing.  Some popular sensors used for magnetic field detection are the flux gate magneto-meters, Hall sensors, and magneto-resistance-based sensors. While temperature
is typically measured with infrared sensors, semiconductor-based sensors, thermo-couples and resistance based sensors. All the sensors mentioned above have the capability to measure either the temperature or the magnetic field, but none can sense both simultaneously.
In recent times, quantum sensors based on NV centers in diamonds have garnered much attention as nano-scale sensors for various physical parameters like magnetic field, temperature, strain, and electric field.
Early observations using NV-based sensors were limited by confocal microscopy, which had a restricted field of view and was not suitable for real-time measurements \cite{maletinsky2012robust}. However, recent advancements in detection techniques, such as the use of photodiode detection and lock-in amplifiers, have overcome these limitations. Photodiode detection provides high sensitivity and fast response times, while lock-in amplifiers improve the signal-to-noise ratio (SNR) by selectively amplifying the desired signal and rejecting noise \cite{acosta2009diamonds}.
In this work, we present a home-built microscope based on photoluminescence (PL) detection of nanodiamonds containing NV$^-$ centers to perform real-time temperature and magnetic field measurements. By employing a lock-in amplifier to improve the SNR, we achieve a mean temperature sensitivity of $\SI{0.4}{\kelvin \per \sqrt{\hertz}}$ and a magnetic field sensitivity of $\SI{3.5}{\mu T \per \sqrt{\hertz}}$.
The ability to simultaneously detect magnetic fields and temperature at the microscale using NV-based nanodiamonds opens new possibilities in various fields. For instance, it can greatly benefit the study of magnetic hyperthermia caused by magnetic nanoparticles in cancer-affected cells, allowing researchers to monitor both local temperature changes and magnetic field distributions in real-time.
This dual-sensing capability of NV-based nanodiamonds, combined with their biocompatibility and high spatial resolution, makes them a powerful tool for advancing our understanding of complex microscale phenomena in biology, physics, and materials science \cite{schirhagl2014nitrogen}. Our work contributes to this growing field by demonstrating a novel approach to real-time, simultaneous dual sensing of temperature and magnetic fields using NV-based nanodiamonds.
\section{Principle of Temperature and Magnetic Field Detection with Nano-diamonds}

The \ce{NV-} centers in nano-diamonds are sensitive to both temperature and magnetic fields due to their electronic structure. The ground state of the \ce{NV-} center is a spin triplet with $m_s = 0$ and $m_s = \pm1 $ sublevels. The energy difference between these sublevels is affected by both temperature and magnetic fields.
For temperature sensing, the zero-field splitting parameter $D$, which determines the energy difference between $m_s = 0$ and $m_s = \pm1$ states, is temperature-dependent is used. By measuring changes in $D$ through optically detected magnetic resonance (ODMR), temperature can be determined with high sensitivity.
However, the magnetic field sensitivity measurements using nanodiamonds are not so straightforward. This is because the sensitivity is affected by several factors such as NV-center concentration, nanodiamond size and preparation method, laser and microwave power, temperature, magnetic field strength and orientation.
The \ce{NV-} center's ground state spin levels ($m_s=0$ and $m_s=\pm1$), as stated earlier are sensitive to external magnetic fields through the Zeeman effect. When an external magnetic field is applied, it causes a splitting of the $m_s=\pm 1$ levels, which can be detected as a shift in the ODMR spectrum. The magnitude of this splitting is proportional to the strength of the magnetic field, allowing for precise magnetic field measurements. The sensitivity of the \ce{NV-} center to magnetic fields can be enhanced by optimizing the ODMR contrast, linewidth, and measurement protocols. The magnetic field value is taken as directly proportional to the difference of the two resonance peaks and the sensitivity calculations are explicitly discussed in the later part of the manuscript.

\section{Experimental Setup}
\noindent We provide a concise overview of the key components and processes involved in the \ce{NV-} center nanodiamond-based sensing experiment. We aim to perform Optically Detected Magnetic Resonance (ODMR) measurements on NV centers in nanodiamonds for real-time detection of thermal and magnetic fields.

\subsection{Optical Setup}

\noindent As shown in Fig.~\ref{fig:schem}, we used a \SI{20}{\micro\liter} solution of nanodiamonds containing 3-ppm NV centers, each with an average diameter of \SI{100}{\nano\meter}. This solution was drop-casted onto a Peltier plate (TEC1-12709). The nanodiamonds were excited with a \SI{532}{\nano\meter} laser, and the resulting photoluminescence (PL) at \SI{637}{\nano\meter} was collected using a 10$\times$, $\num{0.4}$ NA objective. The collected PL was directed to a photodiode (Thorlabs, DET100A2) after passing through filters to block the excitation light.

\subsection{Microwave Setup}

\begin{figure*}
\centering
\includegraphics[width=\textwidth]{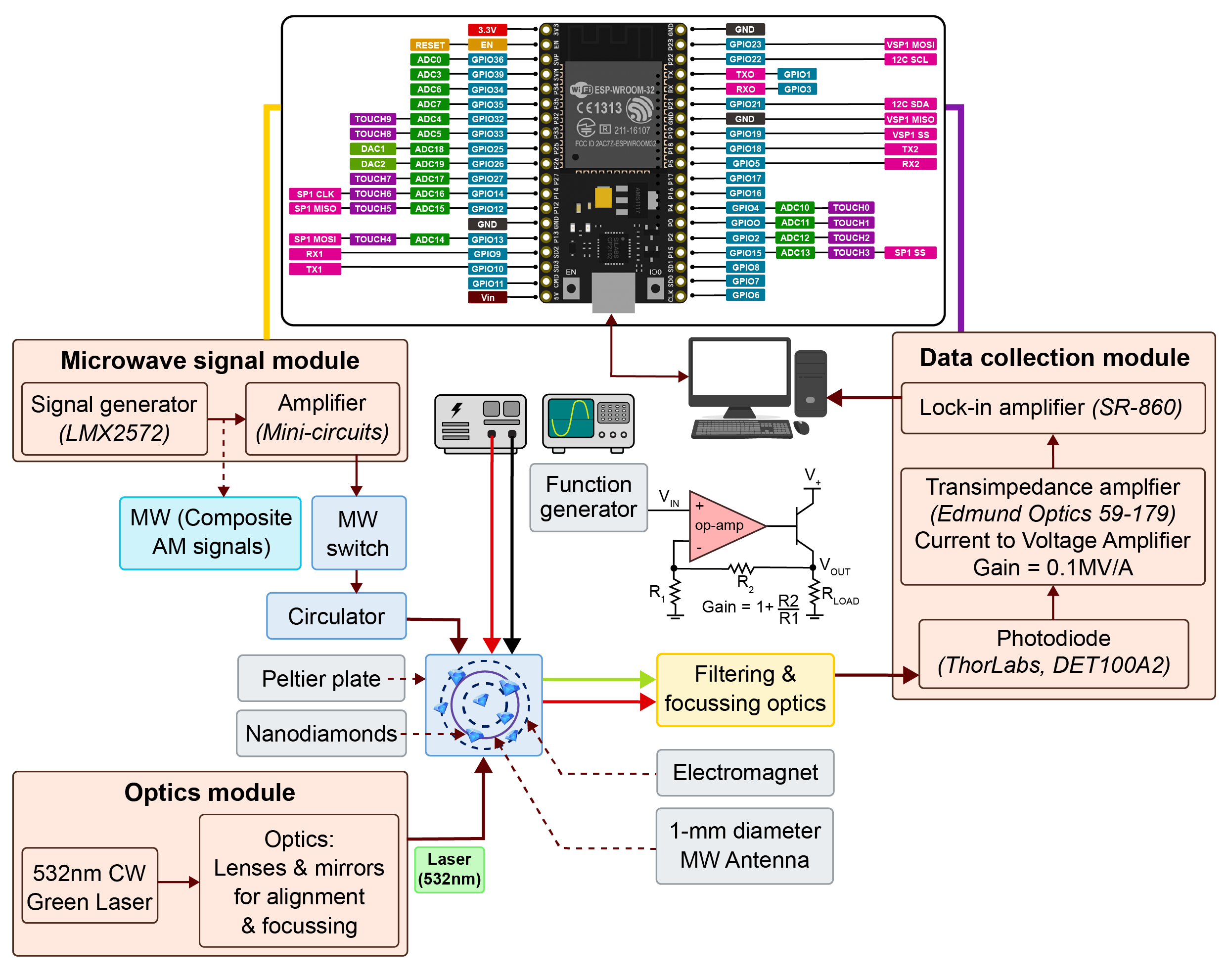}
\caption{(a) Schematic representation of the experimental setup.}
\label{fig:schem}
\end{figure*}

\noindent A microwave loop antenna with a \SI{1.5}{\milli\meter} diameter was mounted on the Peltier plate above the nanodiamond solution (see Fig.~\ref{fig:schem}). The microwave signal was generated by a wideband synthesizer (LMX-2572), amplitude-modulated at \SI{2}{\kilo\hertz}, and amplified (Minicircuits, ZRL-3500+). The LMX-2572 was controlled via an ESP32 microcontroller using the SPI protocol, with default pins: MOSI (GPIO 23), MISO (GPIO 19), SCK (GPIO 18), and CS (GPIO 33).

\subsection{Data Recording}

\noindent A micropython script on the ESP32 controlled the LMX-2572 RF synthesizer, handling power, frequency, and capture settings. During each ODMR cycle (150 total), the script swept frequencies from \SI{2845}{\mega\hertz} in \SI{1}{\mega\hertz} steps. At each step, a trigger signal was sent to a Lock-in Amplifier (Stanford Research Systems, SR860) to ensure synchronized data collection.

\subsection{Calibration of Magnetic Field and Temperature}

\noindent The Peltier plate heated the nanodiamonds, with temperature controlled between \SI{20}{\celsius} and \SI{70}{\celsius}. Two K-type thermocouples measured temperature with \SI{\pm 1.5}{\celsius} accuracy. The NV center, a spin-1 system, has a bright $m_S = 0$ state and darker $m_S = \pm 1$ states. The spin Hamiltonian is:

\begin{equation}
\mathcal{H} = D(T) S_z^2 + \gamma_e \mathbf{B} \cdot \mathbf{S} + E(S_x^2-S_y^2),
\label{eq:hamiltonian}
\end{equation}

where $D(T)$ is the temperature-dependent zero field splitting parameter (\SI{2870}{\mega\hertz}), $\gamma_e$ is the electron gyromagnetic ratio (\SI{28}{\mega\hertz \per \milli\tesla}), $\mathbf{B}$ is the magnetic field, and $E$ is the strain parameter. Without an external field ($\mathbf{B} = 0$), the transition frequencies are:

\begin{equation}
\nu_\pm = D(T) \pm E.
\label{eq:resonance}
\end{equation}

\begin{figure*}[ht!]
\begin{center}
\includegraphics[width=0.8\linewidth]{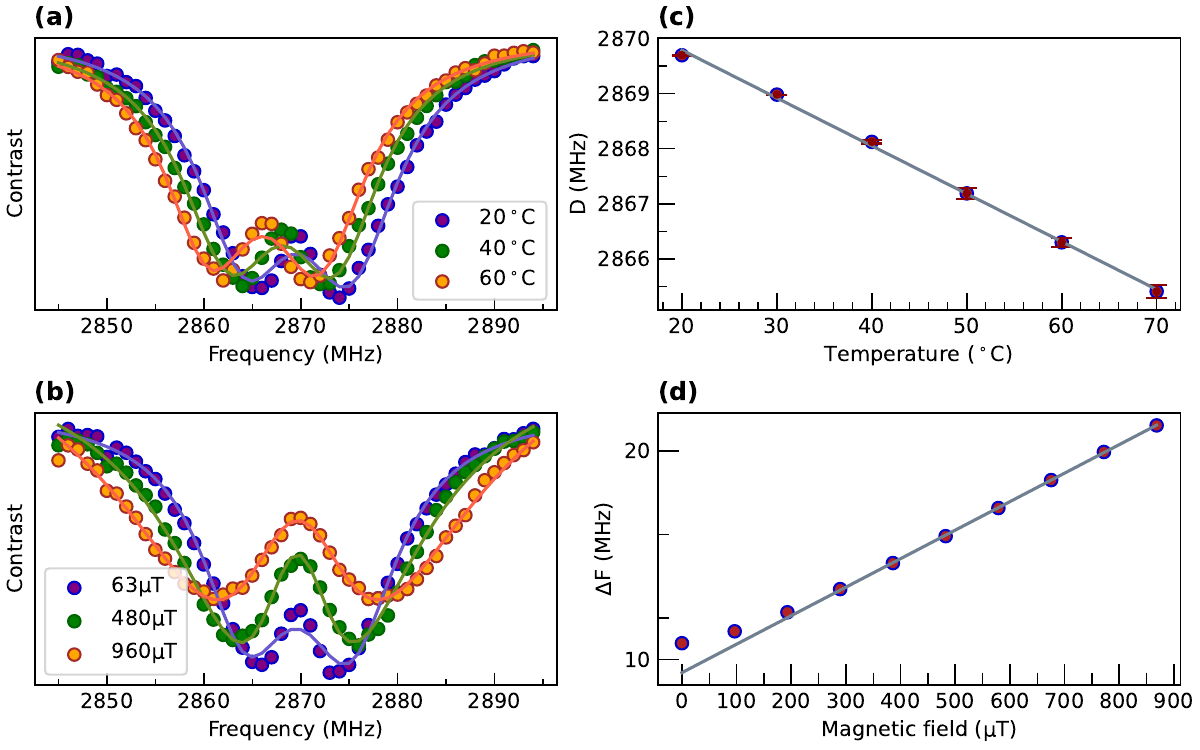}
\caption{(a) Variation of ODMR spectrum with temperature. (b) Variation of ODMR spectrum with magnetic field. (c) The mean shift in zero field parameter $D$ with temperature. (d) The shift in the difference of the resonance frequencies with the variation in magnetic fields.}
\label{fig:DnB}
\end{center}
\end{figure*}

By sweeping the microwave frequency and recording PL, we measured $\nu_\pm$, inferring temperature from Eqn.~\eqref{eq:resonance}. The magnetic field was calculated from the frequency difference. As temperature rises, ODMR plots shift to lower frequencies (Fig.~\ref{fig:DnB}(a)). Magnetic field variations broaden and decrease contrast in the ODMR spectrum (see Fig.~\ref{fig:DnB}(b)). Magnetic field values are derived from the frequency difference, with non-linear behavior initially due to strain. By applying a bias field ($B \gg E$), we achieve a linear response.
\begin{figure}[ht!]
\centering
\centerline{\includegraphics[width=\columnwidth]{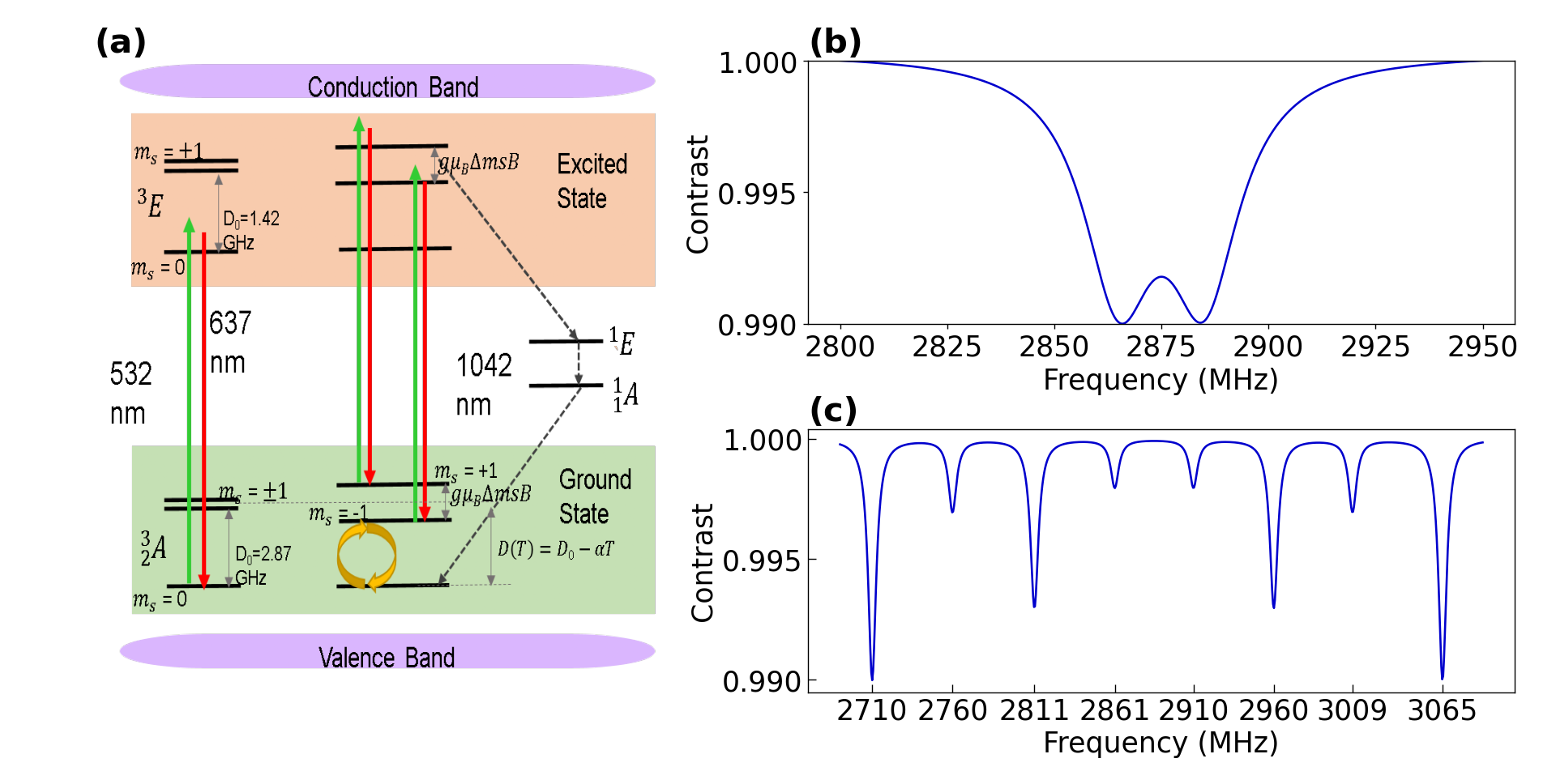}}
\caption{(a) Energy band diagram of NV center based Diamond. (b) ODMR spectrum of NV-based nanodiamonds in the presence of magnetic field. (c) ODMR spectrum of NV-based crystal diamonds in the presence of external magnetic field. The crystal diamond splits into eight distinct peaks due to the crystallographic structure of the diamond which is made up of $4$ axes. The nanodiamonds, on the other hand, do not have any crystalline structure and the NV centers within them have isotropically distributed axes relative to the applied magnetic field and hence exhibit only two peaks in their ODMR spectra, corresponding to the $m_S = 0$ $\to$ $m_S = +1$ and $m_S = 0$ $\to$ $m_S = -1$ transitions, rather than the eight peaks seen in bulk diamond crystals.}
\label{fig:EB dia}
\end{figure}

\section{Results}
\noindent Fig.\,\ref{fig:DnB}(a) shows the measured optically detected magnetic resonance and the corresponding change in ODMR due to changes in temperature in the heat sink and Fig.\,\ref{fig:DnB}(b) shows the measured optically detected magnetic resonance and the corresponding change in ODMR due to changes the magnetic field. Within a given field of view, nanodiamond clusters of different sizes exhibit varying photoluminescence (PL). The temperature and magnetic field generated by the electromagnet are altered to illustrate the shift in the resonance curve. The resonance curves have a mean linewidth of \SI{10}{\mega \hertz}, and the mean strain parameter $E$ is \SI{6.42}{\mega \hertz}. A thorough analysis of the thermal and magnetic field values extracted from the ODMR spectrum, combined with theoretical examination, provides a complete understanding of the system.

It is well known that diamonds form a tetrahedral structure, resulting in four distinct crystallographic directions. When a magnetic field is applied, eight resonances are observed—two for each transition corresponding to the transitions $m_S=0$ to $m_S=\pm1$ along each crystallographic direction—varying in contrast based on the projection of the magnetic field along each of the four directions. However, only two are observed [see Fig.\,\ref{fig:DnB}(a)] for ensembles since the crystal axes of each nanodiamond are randomly oriented, so the NV’s axis can be assumed to be isotropically distributed for many NVs. ODMR data is used to extract the magnetic field. This measures the contrast corresponding to the fluorescence of ground state $m_S = 0$ against mw frequency and holds the information about strain, electric field, magnetic field, and temperature. Yet, the effect of the magnetic field on ensemble NDs is different from that of crystal diamonds. The spectral shape becomes broadened with two dips as the field increases due to Zeeman splitting and the ODMR contrast degrades. The two dips result from lifting the degeneracy of NV’s $|m_s = \pm 1⟩$. The assumption of it being isotropic also makes it insensitive to $B_z$, as there always will be maximally aligned NV within the ensemble.
\subsection{Fitting Model}

The ODMR spectrum shape is approximated as the sum of Lorentzian for each PL dip. The general form of the Lorentzian used is:

\begin{equation}
    L(f_{mw}, f_{\pm}, \delta\nu_{\pm}, C_{\pm}) = \frac{C}{1 + \left(\frac{f_{\pm} - f_{mw}}{\delta\nu_{\pm}}\right)^2}
\end{equation}

where $f_{mw}$ is the applied microwave frequency, $\delta\nu$ is the linewidth, and $C$ is the contrast or the change in fluorescence rate between $|m_s = 0\rangle$ and $|m_s = \pm1\rangle$ state

The zero-field splitting parameter D and the variation of resonance frequencies with changing magnetic field are fit to straight lines:

\begin{equation}
    D = 2871.53 - 0.087T
\end{equation}

\begin{equation}
    \Delta f = 9.37 + 0.014B
\end{equation}

Where T is in $^\circ C$, D $\&$ $\Delta f$ are in \SI{}{\mega \hertz}, and B is in $\mu T$.

\subsection{Sensitivity}

The thermal sensitivity $\eta_T$ is evaluated as:

\begin{equation}
    \eta_T = \frac{\sigma \cdot \sqrt{t_{acq}}}{A \cdot \gamma \cdot \frac{dD}{dT}}
\end{equation}

\begin{figure}[ht!]
\begin{center}
\includegraphics[width=\linewidth]{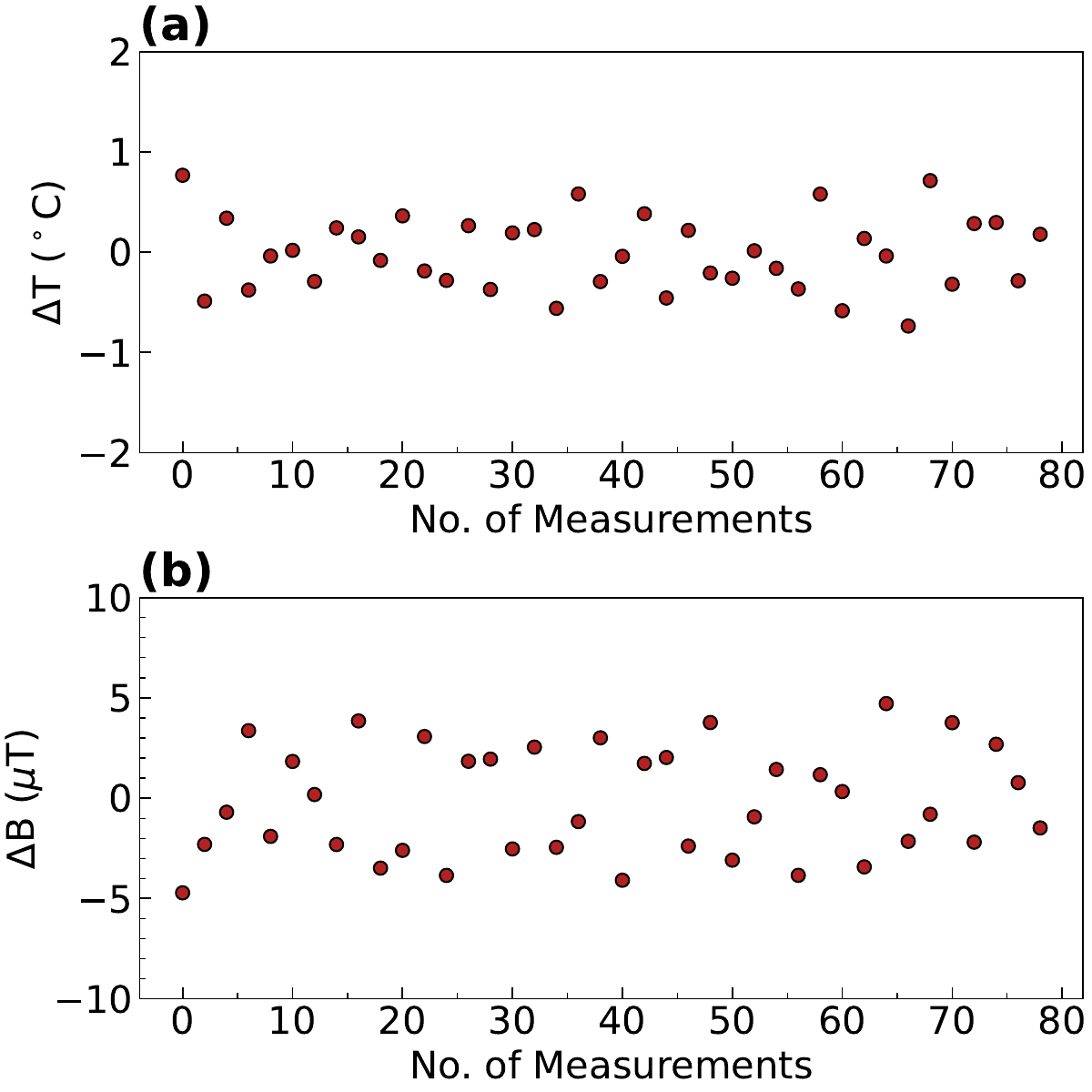}
\caption{(a). Temperature sensitivity standard deviation points appear mostly near the central region. The sensitivity value ranges from \SI{1}{\kelvin \per \sqrt{\hertz}} to the lowest value of \SI{0.4}{\kelvin \per \sqrt{\hertz}} (b) Magnetic field sensitivity standard deviation points appear around the central region but are somewhat distributed. The magnetic field sensitivity ranges from \SI{10}{\mu T \per \sqrt{\hertz}} to the lowest value of \SI{3.5}{\mu T \per \sqrt{\hertz}}}
\label{fig:sens}
\end{center}
\end{figure}

The magnetic field sensitivity $\eta_B$ is evaluated as:

\begin{equation}
    \eta_B = \frac{\sigma \cdot \sqrt{t_{acq}}}{\gamma \cdot \frac{dC}{df}}
\end{equation}

where, $\sigma$ is the standard deviation of the signal when the microwave is kept off, $t_{acq}$ is the signal acquisition time, $A$ is the ODMR contrast, $\gamma$ is the electron gyromagnetic ratio, $\frac{dD}{dT}$ is the temperature dependence of the zero-field splitting, and $\frac{dC}{df}$ is the slope of the ODMR signal, which takes into account the contrast of the ODMR spectrum and the microwave frequency sweep. Fig.\ref{fig:sens} (a) and (b) shows the distribution of temperature and magnetic field sensitivities, with temperature sensitivity ranging from \SI{1}{\kelvin \per \sqrt{\hertz}} to \SI{0.4}{\kelvin \per \sqrt{\hertz}}, and magnetic field sensitivity ranging from $\SI{10}{\mu T \per \sqrt{\hertz}}$ to $\SI{3.5}{\mu T \per \sqrt{\hertz}}$.
\subsection{Simulations}
\begin{figure}[ht!]
    \centering
    \includegraphics[width=\linewidth]{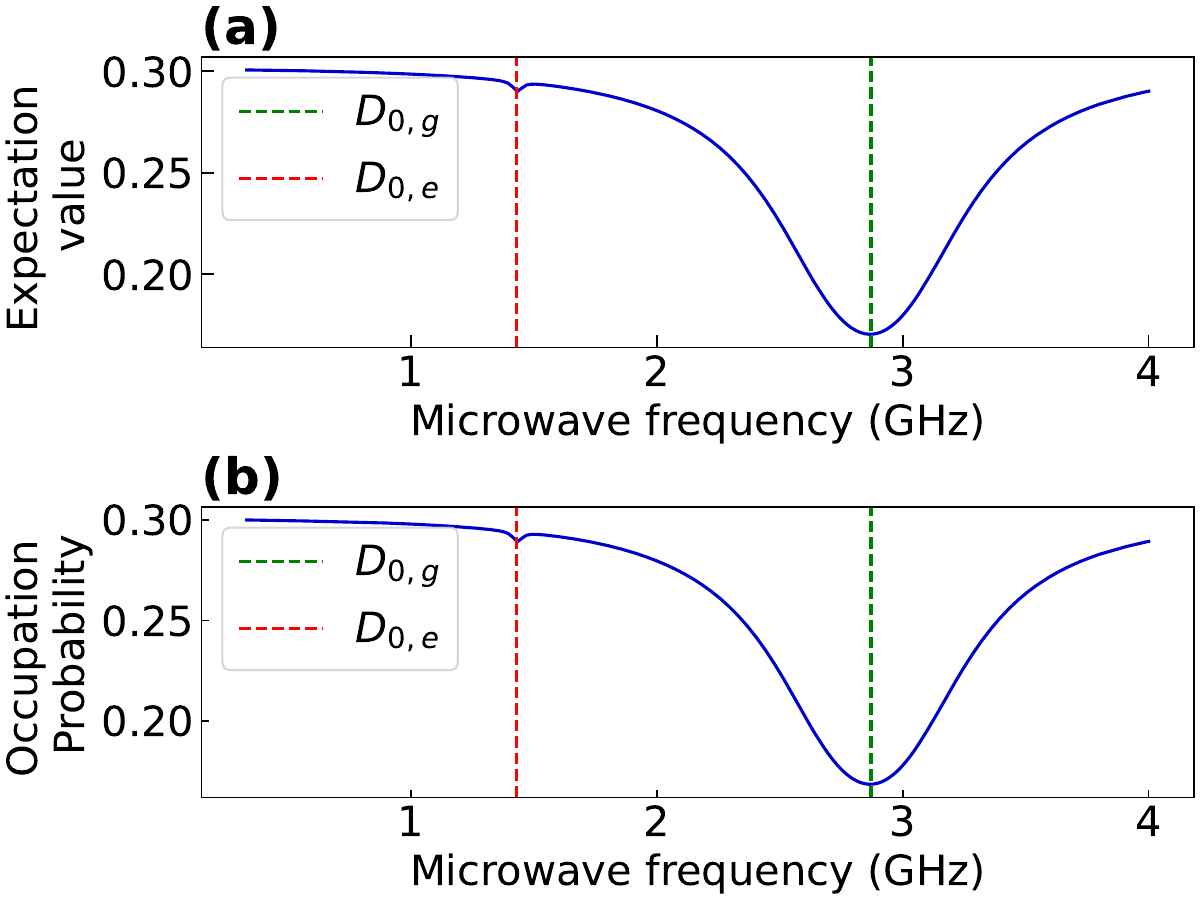}
    \caption{Simulated photoluminescence spectrum.(a) Simulated ODMR using the measurement operator. (b) Simulated ODMR using Lindbald Master Equation for the 7-level system. \\ The dashed line at $D_{0,g} =  \SI{2871.5}{\mega \hertz}$ and $D_{0,e}\SI{1430}{\mega \hertz}$ correspond to the crystal field splitting for the ground state and excited state triplet respectively.The simulations are done at zero magnetic field.}
    \label{fig:sim2}
\end{figure}

In this study, we use simulations based on the Lindblad Master Equation to investigate the dynamics of a seven-level quantum system. Our results demonstrate that the occupancy probability and the temporal evolution of the expectation value of the measurement operator indicate that the ground state probability for the $m_s = 0$ state stabilizes at comparable steady-state values over similar time intervals, as shown in Figure \ref{fig:sim2}. The simulated ODMR spectra, derived from both the measurement operator and the Lindblad Master Equation, provide significant insights into the system's behavior under continuous excitations. Experimentally, the observable is the number of photons emitted due to spontaneous emission, which directly correlates with the population of excited states in the simulation. By weighting the population of these states by their respective decay probabilities, we can predict the photon emission rate. Therefore, the simulation values for excited state populations and decay probabilities quantitatively relate to the experimentally observed photon counts.
\begin{figure}[ht!]
    \centering
    \includegraphics[width=\linewidth]{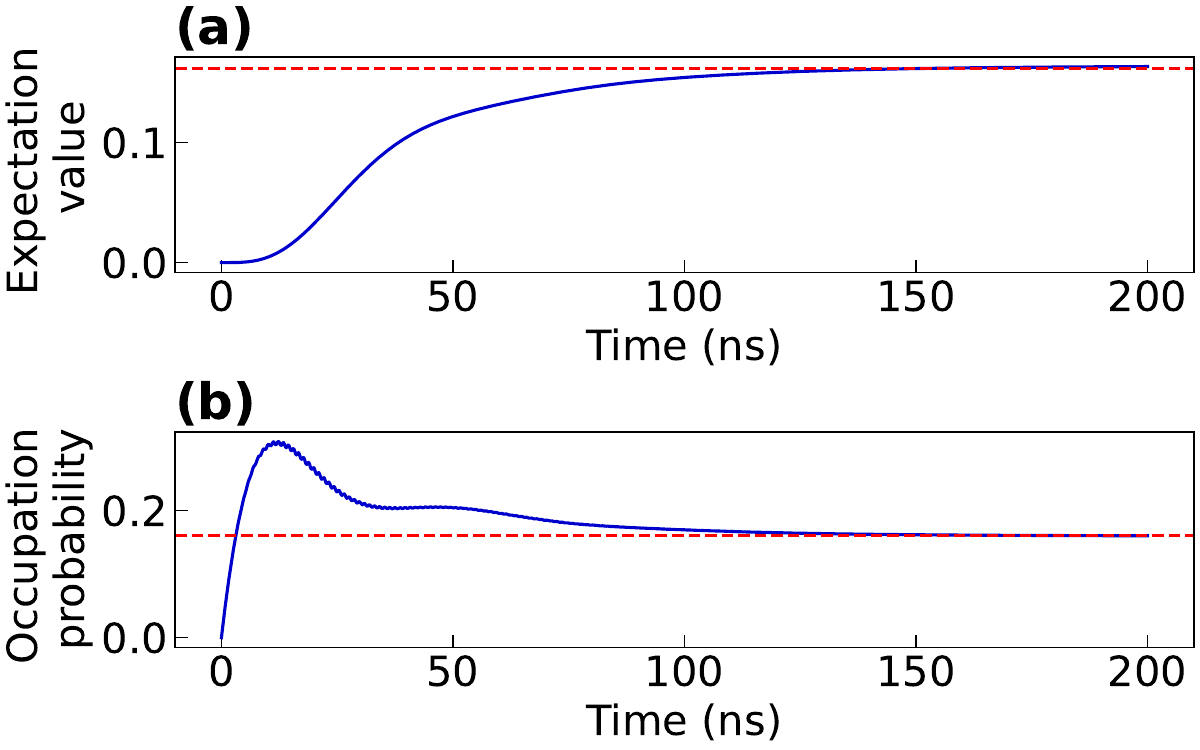}
    \caption{A steady state spin alignment is generated in the triplet ground state under continuous excitation.(a) Time evolution of expectation value of measurement operator. (b) Time evolution of the Occupation probability of $m_s = 0$ ground state. As seen in plots (a) and (b) the ground state  $m_s=0$  probability settle to identical steady state values in similar time intervals. Even though evolution is different, we are only concerned about the steady state.}
    \label{fig:sim1}
\end{figure}
{\small
\begin{equation}
M = k \left( \frac{\gamma_{41}(1+2\epsilon)}{i\sum_i\gamma_{4i}}|4\rangle\langle4| + \\ \frac{\gamma_{51}(1+2\epsilon)}{i\sum_i\gamma_{5i}}|5\rangle\langle5| + \\ \frac{\gamma_{61}(1+2\epsilon)}{i\sum_i\gamma_{6i}}|6\rangle\langle6| \right)
\label{sim_eqn}
\end{equation}}
where the constant k indicates number of photons arriving at the detector and $\epsilon$ represents the ratio of spin non-conserving to spin-conserving transitions. In the simulation, we set k = 1 for simplicity. Also $\epsilon$ is associated with optical pumping and spontaneous emission rates and $\epsilon$ = $k_{24}/ k_{14} = k_{34} / k_{14} = k_{16}/ k_{36} \approx 0.01 $(Table \ref{tab:nv_rates} has the details).
Equation \ref{sim_eqn} defines the measurement operator that counts the photons released during spontaneous transitions, where the constant $k$ represents the number of photons arriving at the detector and $\epsilon$ represents the ratio of spin non-conserving to spin-conserving transitions. For simplicity, we set $k = 1$ in the simulation. The parameter $\epsilon$, associated with optical pumping and spontaneous emission rates, is given by $\epsilon = \frac{k_{24}}{k_{14}} = \frac{k_{34}}{k_{14}} = \frac{k_{16}}{k_{36}} \approx 0.01$ (see Table \ref{tab:nv_rates} for details). The transition rates are determined by various lifetimes and pathways allowed for transitions, such as radiative relaxation, optical pumping, and interstate crossing \cite{masashi_thesis}.

Our simulations reveal that the occupation probability of the ground state peaks at 0.5 for the other spin states and reaches a maximum of 1 for $m_s = 0$. This finding indicates that the Floquet theorem is useful for studying periodic Rabi oscillations. Moreover, the Lindblad Master Equation provides a comprehensive framework for understanding quantum dynamics in the presence of external fields, accurately capturing the system's relaxation and decoherence mechanisms. Although as mentioned the occupation probability and the spin state's fluorescence vary linearly with the corresponding spin population, in simulations, this is achieved by evolving the state and determining the expectation value of the number operator corresponding to the ground state $m_s = 0$ at steady state for each microwave frequency. Consequently, the computation becomes intensive for more frequency points, longer time intervals, and the presence of external fields. Our simulation results, shown in Figure \ref{fig:sim2} (a) and (b), are based on a single spin in the NV center in a diamond. We believe that more precise sensitivity analysis could be achieved by studying an ensemble of NV centers in nanodiamonds, providing more detailed insights than that already available in literature. Additionally, future experimental advancements could focus on using pulsed lasers and applying the microwave field within those pulses.
\begin{table}[ht]
%\footnotesize
%\centering
\setlength{\tabcolsep}{3pt}
\begin{tabular}{|p{3cm}|p{2cm}|p{2cm}|}
    \hline
    Description & Symbol & Value (MHz)\\
    \hline
    Optical Pumping rate & $k_{14} = k_{25} = k_{36}$ & 64 MHz \\
    \hline
    Spontaneous Emission Rate & $k_{41} = k_{52} = k_{63}$ & 64 MHz \\
    \hline
    Interstate crossing rates from $^3E$ to singlet & $k_{57} = k_{67}$ & 79.8 MHz \\
    & $k_{47}$ & 11.8 MHz \\
    \hline
    Interstate crossing rates from singlet to $^3A_2$ & $k_{72} = k_{73}$ & 0 MHz \\
    & $k_{71}$ & 5.6 MHz \\
    \hline
\end{tabular}
%\title{displays the transition rates considered for the simulations}
\caption{displays the transition rates considered for the simulations}
\label{tab:nv_rates}
\end{table}
The values mentioned in table \ref{tab:nv_rates}are the average of the values taken from: \cite{masashi_thesis}, \cite{mw_cavity_cooling}, \cite{NVstructure}, and \cite{optical_magnetic_imaging}

\begin{figure*}[ht!]
\begin{center}
\includegraphics[width=\linewidth]{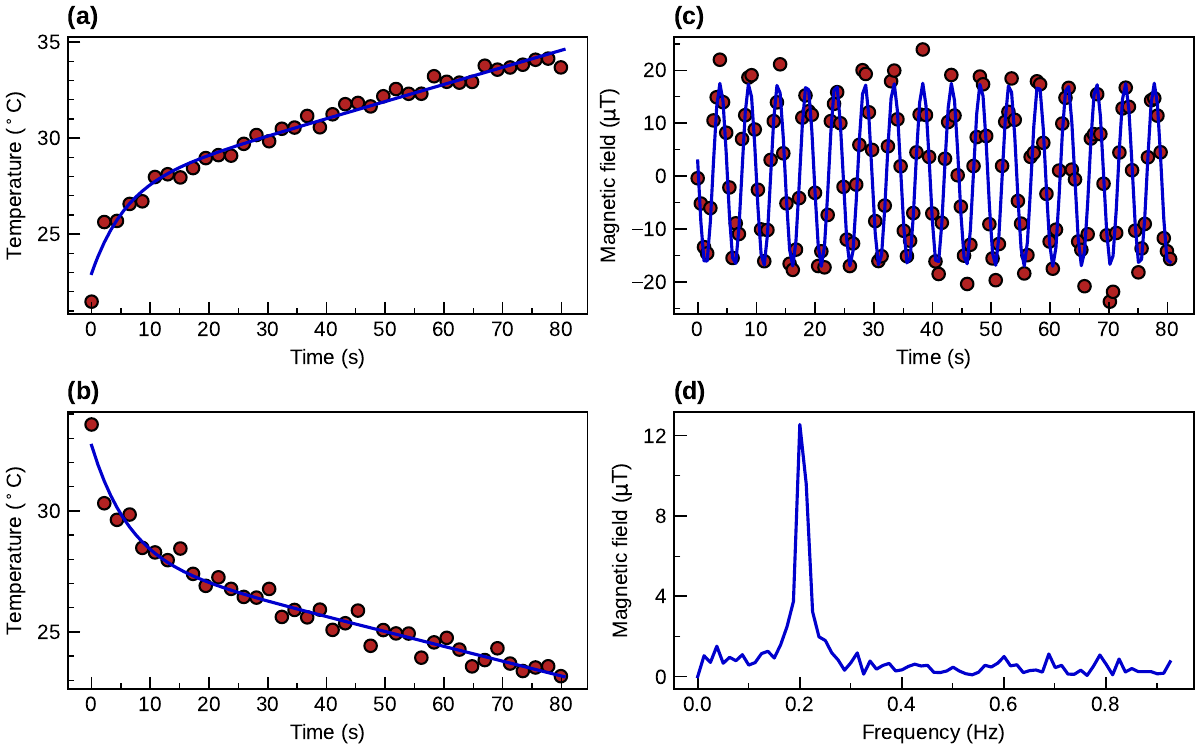}
\caption{(a). Rise Time calculation plot using NV center Nanodiamonds (b). Fall Time calculation plot using NV center Nanodiamonds (c). Simultaneous magnetic field and temperature detection using a sinusoidal signal with a time-period of $\SI{5}{\milli s}$ (d). Fourier domain representation of plot (c) showing the existence of the signal at $\SI{200}{\milli \hertz}$ and no other harmonic exists.}
\label{fig:temporal}
\end{center}
\end{figure*}

We next show the temporal measurement in Fig.\,\ref{fig:temporal}. In Fig.\,\ref{fig:temporal}(a) and (b), the
temporal profile of the rise time and fall time of the temperature at the location of the nanodiamonds on the Peltier plate is shown. The temperature rise and fall follow an exponential increase and decrease curve and in \SI{1.5}{\minute} the temperature changes from room temperature to \SI{34}{\textcelsius} and vice-versa. The plots are fitted using exponential incremental and decremental curves and the exponent value gives us the rise time and fall time values respectively. As we measured the rise and fall times we had magnetized the electromagnet to generate the required magnetic field which we desired to measure using the NVND-based sensor. For the magnetic field determination sinusoidal signal is sent to the electro-magnet which in return generates the magnetic field experienced by the nanodiamonds. The magnetic field is measured as a function of the difference in the resonant frequencies at the location of the nanodiamonds.
When the \ce{NV-} centers in the nanodiamond are out of resonance with the microwave signal, there is no
modulation in the PL. However, when the the microwave frequency is close to the resonance, there is a large
modulation in the PL which is recorded by the lockin camera.  To get the temporal profile of the nanodiamonds, we turn on the transistor. As the transistor heats up, we record the ODMR amplitude using the above mentioned technique by recording a frame with an integration time of \SI{5}{\second}.

In this work we employ a range of microwave frequency to detect the ODMR amplitude variation and infer the temperature and magnetic field values. By tracking the entire ODMR spectrum, we estimated the magnetic field
information simultaneously with the temperature variation by fitting the Hamiltonian parameters appropriately. However, tracing out the full microwave spectra is time consuming and detrimental for real time
detection. A potential methodology would be to perform multi point ODMR on a widefield setup\,\cite{nishimura2021wide,PhysRevResearch.2.043415}. Further, new methods need to be developed for simultaneous real time tracking of magnetic and thermal variations using NV centers in diamond.

\section*{Conclusion}
In this work, we have developed a novel magnetic field and thermal sensor that detects the temperature and magnetic field simultaneously in real-time with a mean thermal sensitivity of \SI{0.4}{\kelvin \per \sqrt{\hertz}} and mean magnetic field sensitivity of $\SI{3.5}{\mu T \per \sqrt{\hertz}}$ using \ce{NV-} centers in nanodiamonds.
We tested the magnetic field and thermal detection of nanodiamonds mounted on a Peltier plate. In the future, we will extend the current method to improve on the recorded sensitivity numbers by improving the readout techniques and integrating them with microelectronic circuits, and developing on-chip solutions could pave the way for widespread adoption in consumer electronics, biomedical devices, and portable diagnostic tools.

\section*{Author Contribution}
\noindent S.S and D.S conceived the idea. S.S designed the experimental setup, wrote the software control for the experiments, performed the experiments and along with D.S analyzed the data. S.S and D.S wrote the manuscript. N.M did the simulations. All authors reviewed and approved of the manuscript. K.S supervised all aspects of the work.

\section*{Acknowledgment}
\noindent K.S. acknowledges funding from AOARD Grant No. FA2386-23-1-4012 and DST-SERB Power Grant No. SPG/2023/000063. S.S. acknowledges funding support from Meity NNetra project. The authors acknowledge helpful discussions with Prof. Himadri Shekhar Dhar and Prof. Pradeep Sarin.

%\tableofcontents

%\bibliographystyle{apsrev4-1}
%\bibliography{references}


\end{document}